\documentstyle[amstex,eqsecnum,aps,multicol,epsfig]{revtex}

\begin{document}
\title{Entangled graphs: Bipartite entanglement in multi-qubit systems}
\author{Martin Plesch${}^{1}$ and  Vladim\'{\i}r Bu\v{z}ek${}^{1,2}$}
\address{
${}^{1}$Research Center for Quantum Information, 
Slovak Academy of Sciences, D\'{u}bravsk\'{a} cesta 9, Bratislava, Slovakia
\\
${}^{2}$Department of Mathematical Physics, National University of Ireland,
Maynooth, Co. Kildare, Ireland
}
\date{November 2, 2002}
\maketitle

\begin{abstract}
Quantum entanglement  in multipartite systems
cannot be shared freely. In order to illuminate 
basic rules of entanglement sharing between qubits
we introduce a concept of an  
entangled structure (graph) such that  each qubit
of a multipartite system
is associated with a point (vertex)  while a bi-partite
entanglement between two specific qubits 
is represented by a connection (edge) between these points. 
We prove that any such entangled structure can be associated 
 with a {\em pure} state of a  multi-qubit system.
Moreover, we show that a pure state corresponding to a given
entangled structure is a superposition of vectors  from a subspace 
of the $2^N$-dimensional Hilbert space, whose dimension grows {\em linearly}
with the number of entangled pairs.
\end{abstract}

\begin{multicols}{2}
\section{Introduction} The
entanglement is a key ingredient of quantum mechanics 
\cite{Schroedinger,Peres}. In the last decade it has been
identified  as a key resource for
quantum information processing. In particular,
quantum computation \cite{Preskill,Nielsen},
quantum teleportation \cite{Bennett93}, quantum dense coding
\cite{Bennett92},  certain types of quantum key distributions
\cite{Ekert91} and quantum secret sharing protocols \cite{Hillery99},
are based on the existence of entangled states.

The nature of quantum entanglement
between two qubits is well understood by now. In particular, the
necessary and sufficient condition for inseparability of two-qubit
systems has been derived by Peres \cite{Peres96} and Horodecki et al.
\cite{Horodecki96}. Reliable measures of bi-partite entanglement
have been introduced 
and well analyzed (see for instance Refs.~\cite{Vedral97,Horodecki2001}).
On the other hand 
it is a very difficult task to generalize the analysis of entanglement from
two to multi-partite systems. 
The multi-partite entanglement is a complex phenomenon.
One of the reasons is that
 quantum entanglement cannot be shared  freely among many
particles. 
For instance, having four qubits, we are able to prepare a state with two
e-bits (two Bell pairs, as an example), but not more. This means that 
the 
structure of quantum mechanics imposes strict bounds on bi-partite
entanglement in multi-partite systems. This issue has been first addressed
by Wootters et al. \cite{CKW,Chains} who have derived important bounds
on shared bi-partite entanglement in multi-qubit systems. In fact, one
can solve a variational problem to answer a question:
What is a
pure multi-partite state with specific constraints on bi-partite
entanglement? O'Connors and Wootters \cite{Chains} have studied what is
the state of a multi-qubit ring with maximal possible entanglement between
neighboring qubits. Another version of the same problem has been analyzed
by Koashi et al. \cite{Koashi} who have derived an explicit expression for
the multi-qubit completely symmetric state (entangled web)
in which all possible pairs of qubits are maximally entangled. 

In his recent work D\"{u}r \cite{Dur} has introduced a concept of 
{\em entanglement molecules}
D\"{u}r has shown that an arbitrary entanglement
molecule can be represented by a {\em mixed} state of a multi-qubit system.
On the other hand in his work the problem of pure multi-partite states with
specific  entangled pairs of qubits has not been discussed thoroughly.
Specifically,   D\"{u}r has considered just the condition of inseparability 
for given set of pairs, but he did not impose a strict condition
of separability for the remaining pairs of qubits.

Following these ideas we analyze in the present paper an object, entangled
structure, which will be called through the paper  as the {\em entangled
graph} \cite{note}. In the graph, each qubit is represented as a vertex and
an edge between two vertices denotes entanglement between these two
particles (specifically, the corresponding two-qubit density operator is
inseparable). The central issue of the paper is to show that any entangled
graph with $N$ vertices and $k$ edges can be associated with a {\em pure}
multi-qubit state. We prove this result constructively, by showing the
explicit expression of corresponding pure states. We show that any entangled
graph of $N$ qubits can be represented by a pure state from a subspace of
the whole
$2^N$-dimensional Hilbert space of $N$ qubits. The dimension of this 
subspace is at most quadratic in  number of qubits.

\section{Simple example}

In general an $N$-partite system can exhibit various types of multi-partite
correlations, ranging from bi-partite entanglement to intrinsic
multi-partite correlations of the GHZ nature. 
Correlations associated with the system specify its state (certainly,
this specification is not necessarily unique). Ideally we would like to know
the whole hierarchy of quantum correlations in the multi-partite system.
We are able to determine and quantify bi-partite quantum
correlations. Unfortunately,   for existence of 
intrinsic $N$-qubit correlations we even do not
have sufficient and necessary conditions (see
Refs.~\cite{Horodecki2001,CKW}). 
 Nevertheless, as
suggested by Coffman, Kundu and  Wootters  \cite{CKW} 
it is very instructive to
understand how a bi-partite entanglement is ``distributed'' in $N$-qubit
system. The inequalities derived by these authors (the so-called CKW
inequalities) open new possibilities how to understand the complex
problem of bounds on shared entanglement.  
The CKW inequalities utilize the measure of entanglement called concurrence
as introduced by Wootters et al. \cite{Konkurencia}. This measure is defined
as follows: Let us assume a two-qubit system prepared in the state
described by the density operator $\rho$.
From this operator one can evaluate the so-called 
spin-flipped operator defined as
\begin{equation}
\tilde{\rho}=(\sigma _{y}\otimes \sigma _{y})\rho ^{\ast }(\sigma
_{y}\otimes \sigma _{y}),
\end{equation}
where $\sigma _{y}$ is the Pauli matrix and a star denotes a complex
conjugation. Now we define the matrix 
\begin{equation}
R=\rho \tilde{\rho}\, ,
\label{R}
\end{equation}
and label its (non-negative) eigenvalues, 
in decreasing order $\lambda _{1},\lambda _{2,}\lambda
_{3}$ and $\lambda _{4}$. The concurrence is then defined as 
\begin{equation}
C={\rm max}\left\{ 0,
\sqrt{\lambda _{1}}-\sqrt{\lambda _{2}}-\sqrt{\lambda _{3}}-%
\sqrt{\lambda _{4}}\right\}  \label{C}\, .
\end{equation}
This function serves as an indicator whether the two-qubit system is
separable (in this case $C=0$), while for $C>0$  it  
measures the amount of bipartite entanglement between two qubits
with a number between $0$ and $1$. Larger the value of $C$  stronger
the entanglement between two qubits is.

Unfortunately, no simple measures of entanglement are known for
multi-qubit systems. Nevertheless it is still of importance to
understand how a bi-partite entanglement is distributed in
$N$ qubit system. In this paper we will utilize a concept of entangled graph
to illuminate some aspects of the problem.

Using the concurrence 
we can easily associate an entangled graph with every $N$-partite state.
On the other hand
there is no one to one correspondence between graphs and states. 
For instance all separable states with $N$
qubits have the same graph -- $N\,$\ vertices and no edges. 
Also all $GHZ$-like states
for $N>2$ would have the same graph. The question we are going to address
can be formulated as follows:
Is 
it possible to construct at least one {\em pure} state for a given graph? 

\begin{figure}[tbph]
\centerline {\epsfig{width=8.0cm,file=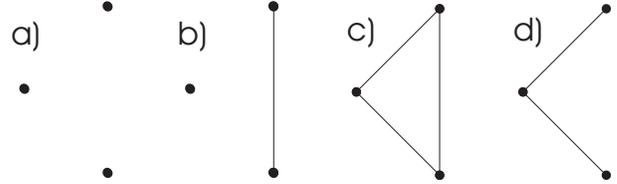}}
\bigskip
\caption{
Four different classes of entangled 
graphs associated with states of three
qubits:
{\bf  a)} separable and GHZ-like states with no bi-partite entanglement, 
{\bf b)} 
Bell-like states, {\bf c)} $W$-like states and {\bf d)} a new category
of entangled states
}
\label{3ent}
\end{figure}

We start our discussion with the simplest non-trivial example:
 Let us consider three qubits.
Pure states of three qubits  can be divided into six classes 
(see Ref.~\cite{3qubit}):
separable states, bipartite entangled states (three classes respective to the
permutation), the $W$-type states and the $GHZ$-type states. 
On the other hand
in Fig.~\ref{3ent} we represent 
all possible graphs for a three-partite system. Separable
and $GHZ$-like states correspond to the case ({\em a}), 
bipartite entangled states
are represented by the graph  ({\em b}),
while the $W$-type states are represented by the graph ({\em c}). 
Obviously, one can imagine also an additional type of a graph,
when a given qubit (labelled as the qubit 2) 
is entangled with two others (labelled as 1 and 3, respectively), while
the qubits 1 and 3 are not entangled. The question is whether this type of
a graph (see Fig.~1d)
can exist. Does not the entanglement between qubits 1 and 2, and
2 and 3 induce the entanglement between qubits 1 and 3?

In order to illuminate this simple problem let us first consider
mixed states. According to D\"{u}r (see
Ref.~\cite{Dur})  {\em mixed} states associated
with the graph $d)$ do exist and have the form
\begin{eqnarray}
\rho =a|\Psi ^{+}\rangle _{12}\langle \Psi ^{+}|
\otimes
|0\rangle _{3}\langle 0|
+  (1-a)|\Psi ^{+}\rangle _{23}\langle \Psi ^{+}|
\otimes
|0\rangle _{1}\langle 0|\, , 
\label {2.4}
\end{eqnarray}
where 
$|\Psi ^{+}\rangle=\left( |01\rangle +|10\rangle \right)/\sqrt{2}$ is a Bell
state. 
On the other hand 
by  inspection one can check that for 
$0<a<1$ the mixed state (\ref{2.4}) exhibits required correlations.

From the explicit expression for the mixed state (\ref{2.4}) associated
with the graph $d)$ we might try to express a {\em pure} state
corresponding to the same graph as follows
\begin{equation}
|\Phi \rangle =\alpha |\Psi ^{+}\rangle _{12}|0\rangle _{3}+\sqrt{1-\alpha
^{2}}|\Psi ^{+}\rangle _{23}|0\rangle _{1}.
\end{equation}
This state has requeired properties imposed on correlations betweem qubits
1-2 and 1-3. But, it
also exhibits entanglement between qubits 2 and 3,
and therefore the corresponding graph is of the type $c)$.
In order to find a pure state for the graph $d)$ we will consider a whole
family of states of the form
\begin{equation}
|\Phi \rangle =\alpha |000\rangle +\beta |100\rangle +\gamma |110\rangle +%
\sqrt{1-\alpha ^{2}-\beta ^{2}-\gamma ^{2}}|111\rangle .  \label{3retiazka}
\end{equation}
It is straightforward to calculate concurrencies for all pairs of qubits
in the state (\ref{3retiazka}).  
We find  that $C(1,3)$ 
is always zero while $C(1,2)$ and $C(2,3)$ are non-zero
for all values of involved probability amplitudes.
By inspection it is possible to determine that the state 
(\ref{3retiazka}) belongs to the class of GHZ states,
since it contains intrinsic three-partite entanglement.

We can conclude this simple example by saying that 
all 3-qubit entangled graphs can be realized by pure
3-qubit states.
We note that the classification of states 
according to entangled graphs, representing the two-partite
entanglement is incompatible with the classification presented in  
Ref.~\cite{3qubit}. We see that two types of states 
(GHZ states and separable states) have the same graph. On the other hand
two states of the same class (GHZ states) can be represented by different
graphs.

\section{N-particle system}

Let us first consider entangled graphs associated with mixed 
$N$-qubit states.
These graphs  consist of $N$ vertices. 
Let the parameter $k$ denote the
number of edges in the graph, with the condition 
\begin{equation}
0\leq k\leq \frac{N(N-1)}{2}.
\end{equation}
Then let us define a set $S$ with $k$ members. These will be pairs of
qubits between which we expect entanglement; thus for every $i<j$
\begin{eqnarray}
\label{3.2}
\{i,j\} &\in &S\qquad \Longleftrightarrow \qquad C(i,j)>0 \\
\{i,j\} &\notin &S\qquad \Longleftrightarrow \qquad C(i,j)=0.  \nonumber
\end{eqnarray}
A state of the form 
\begin{equation}
|\Psi \rangle _{ij}=|\Psi ^{+}\rangle _{ij}|0...0\rangle _{\overline{ij}}
\label{3.3}
\end{equation}
exhibits 
entanglement between qubits $i$ and $j$ and nowhere else. In Eq.~(\ref{3.3})
the vector 
$|\Psi^{+}\rangle _{ij}=\left( |01\rangle +|10\rangle \right)/\sqrt{2} $  
represents the
maximally entangled Bell state between qubits of two qubits $i$ and $j$.
The rest of $N-2$ qubits are assumed to be in the product state 
$|0...0\rangle_{\overline{ij}}$.
D\"{u}r in Ref.~\cite{Dur} has proposed
a {\em mixed}
state of $N$ qubits, which corresponds to a graph defined by the set $S$
in the form
\begin{equation}
\rho =\frac{1}{k}\sum_{\{i,j\}\in S}|\Psi \rangle _{ij}\langle \Psi |_{ij}.
\label{3.4}
\end{equation}
It is much more complex task to find a {\em pure} state of $N$ qubits
corresponding to a specific graph. We will solve this problem below.

\subsection{Pure states}
\begin{figure}[tbph]
\centerline {\epsfig{width=8.0cm,file=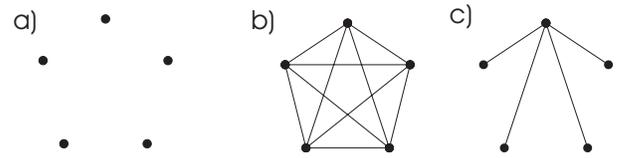}}
\bigskip
\caption{Examples of entangled graphs associated with states of
5 qubits: {\bf a)} separable states, or any
other states with no bipartite entanglement; {\bf b)} $W$-type states;
{\bf c)} star-shaped states }
\label{5ent}
\end{figure}

We start our analysis with entangled graphs that exhibit specific
symmetries. Certainly the two most symmetric graphs are
those representing separable states [no edges 
 - see Fig.\ref{5ent}~a] and those representing $W$-states, with
all vertices connected by edges. A representative of a pure completely
separable state is described by the vector
$|\Psi \rangle =|0...0\rangle $. 
The 
W-state $|W\rangle _{N}=1/\sqrt{N}|N-1,1\rangle $, is a maximally symmetric
state with  one qubit in state $|1\rangle$ 
and $N-1$ qubits in state $|0\rangle$ 
(see Refs.~\cite{Koashi,Dur}). This state
maximizes the bi-partite concurrence - its value is given by the expression
$C=2/\sqrt{N}$. We see that the most symmetric entangled graphs 
do correspond to specific pure multi-qubit states.

Let us now consider graphs with a lower symmetry. 
For instance, a
star-shaped graph (Fig.\ref{5ent}.~c). In this case, the given 
qubit is (equally)
entangled with all other qubits in the system, that in turn are not
entangled with any other qubit. 

D\"{u}r \cite{Dur} 
has proposed an explicit expression for 
a pure state associated with 
this type of entangled graph:
\begin{equation}
|\Psi \rangle =\frac{1}{\sqrt{2}}|1\rangle |0...0\rangle +\frac{1}{\sqrt{2}}%
|0\rangle |N-2,1\rangle \, .
\label{3.5}
\end{equation}
In fact, this state maximizes the concurrence
between the first and any other qubit. But, the remaining qubits are still
mutually entangled. So the state (\ref{3.5}) is represented by the graph
$b)$ (all vertices are connected) rather than graph $c)$.
In our analysis we require more stringent constraints than in
Ref.~\cite{Dur}, where only the conditions on the 
presence of entanglement
between specific qubits have been imposed. We require 
the conditions (\ref{3.2}), that is the presence and absence of bi-partite
entanglement for given pairs of qubits in the system.

We find that 
a pure state which indeed is represented by the star-shaped graph (see
Fig.~\ref{5ent}~c) is given by the expression
\begin{equation}
|\Psi \rangle =\alpha |W\rangle _{N}+\beta |0\rangle |1...1\rangle 
\label{3.6}
\end{equation}
with the normalization 
condition $\left| \alpha \right| ^{2}+\left| \beta \right|
^{2}=1$. 
For $N>4$, the reduced two-qubit density operator for the first
and any other qubit in the system reads 
\begin{equation}
\rho _{1i}=\left( 
\begin{array}{cccc}
\frac{N-2}{N}\left| \alpha \right| ^{2} & 0 & 0 & 0 \\ 
0 & \left| \alpha \right| ^{2}\frac{1}{N}+\left| \beta \right| ^{2} & \left|
\alpha \right| ^{2}\frac{1}{N} & 0 \\ 
0 & \left| \alpha \right| ^{2}\frac{1}{N} & \left| \alpha \right| ^{2}\frac{1%
}{N} & 0 \\ 
0 & 0 & 0 & 0
\end{array}
\right) .
\label{1}
\end{equation}
One of the eigenvalues of the partially transformed matrix 
\begin{equation}
\lambda =\left| \alpha \right| ^{2}\frac{n-2-\sqrt{n^{2}+8-4n}}{n}.
\end{equation}
is negative for every $\alpha >0$. Consequently, using the Peres-Horodecki
criterion we see that the first qubit is indeed entangled with any other
qubit in the system
for any non-trivial value of $\alpha 
$. Now we have to show, that all other qubits in the system are not 
mutually entangled (i.e.
all pairs of qubits $\{i,j\}$, where $1<i<j<N$ are separable). The reduced
density operator describing a state of qubits $i$ and $j$ reads 
\begin{equation}
\rho _{ij}=\left( 
\begin{array}{cccc}
\frac{N-2}{N}\left| \alpha \right| ^{2} & 0 & 0 & 0 \\ 
0 & \left| \alpha \right| ^{2}\frac{1}{N} & \left| \alpha \right| ^{2}\frac{1%
}{N} & 0 \\ 
0 & \left| \alpha \right| ^{2}\frac{1}{N} & \left| \alpha \right| ^{2}\frac{1%
}{N} & 0 \\ 
0 & 0 & 0 & \left| \beta \right| ^{2}\, .
\end{array}
\right) 
\label{2}
\end{equation}
The smallest eigenvalue of the partially transposed operator is
\begin{equation}
\lambda  =\frac{N-2\left| \,\alpha \right| ^{2}-\sqrt{\delta }}{2\,N} 
\end{equation}
where
\begin{equation}
\delta  =N^{2}-4\,\left| \alpha \right| ^{2}\,\left( N-1\right)
\,N+4\,\left| \alpha \right| ^{4}\,\left( 2+\left( N-2\right) \,N\right) . 
\nonumber
\end{equation}
We see that 
for all $\alpha$ such that
\begin{equation}
\left| \alpha \right| \leq \frac{\sqrt{N^{2}-2N}}{N-1}.
\end{equation}
the smallest eigenvalue $\lambda$ is non-negative. Consequently,
the corresponding density operator is separable.
Thus we have found a family of states, that correspond to 
 the desired graph. For
the special case of $N=4$, the reduced operators have a form different 
from (\ref{1}) and (\ref{2}) and
also the final condition is more complicated. However, it is quite easy to
find one example of the state of four qubits corresponding to the
star-shaped graph. The state vector reads:
\begin{eqnarray}
|\Psi \rangle  &=&\frac{1}{\sqrt{5}}\left( |0111\rangle +|0001\rangle
+|0010\rangle +|0100\rangle +|1000\rangle \right)
\nonumber \\
&=&\frac{2}{\sqrt{5}}|W\rangle _{4}+\frac{1}{\sqrt{5}}|0111\rangle . 
\end{eqnarray}

Above we have analyzed the most symmetric entangled graphs. In what follows
we propose a 
general algorithm how to construct a pure state for an arbitrary graph. 
Let us
consider a pure state  of $N$ ($N>4$) qubits  
described by the vector 
\begin{equation}
\label{3.12}
|\Psi \rangle =\alpha |0..0\rangle +\beta |1..1\rangle +\sum_{\{i,j\}\in S}%
\frac{\gamma }{\sqrt{k}}|1\rangle _{i}|1\rangle _{j}
|0...0\rangle_{\overline{ij}}
\end{equation}
with the normalization condition 
$\left| \alpha \right| ^{2}+\left| \beta \right|
^{2}+\left| \gamma \right| ^{2}=1$. 
In what follows we will show  that for a certain range
of parameters this state matches a graph given by the condition
(\ref{3.2}).

Firstly,  we 
show that a pair of qubits $i$ and $j$ such that 
$%
\{i,j\}\in S$ is indeed entangled. 
The corresponding reduced density operator reads
\begin{eqnarray}
&&\rho _{ij}=
\label{3.14}
\\
&&\left( 
\begin{array}{cccc}
\left| \alpha \right| ^{2}+\left| \gamma \right| ^{2}\frac{k-n_{i}-n_{j}+1}{k%
} & 0 & 0 & \frac{\alpha \gamma ^{\ast }}{\sqrt{k}} \\ 
0 & \left| \gamma \right| ^{2}\frac{n_{i}-1}{k} & \left| \gamma \right| ^{2}%
\frac{n_{ij}}{k} & 0 \\ 
0 & \left| \gamma \right| ^{2}\frac{n_{ij}}{k} & \left| \gamma \right| ^{2}%
\frac{n_{j}-1}{k} & 0 \\ 
\frac{\alpha ^{\ast }\gamma }{\sqrt{k}} & 0 & 0 & \left| \beta \right| ^{2}+%
\frac{\left| \gamma \right| ^{2}}{k}
\end{array}
\right)
\nonumber
\end{eqnarray}
where $n_{i}$ is the number of connections
originating from the $i$th vertex (the number of qubits we wish to have
entangled with the $i$th one) and $n_{ij}$ is the number of vertices, 
that are
connected directly with the $i$th and $j$th vertex. The following
inequalities for these variables hold: 
\begin{eqnarray}
1 &\leq &n_{i}\leq k  \label{ni}\, ; \nonumber
\\
0 &\leq &n_{ij}<\frac{k}{2}\, ;  \label{nij} 
\nonumber
\\
2 &\leq &n_{i}+n_{j}\leq k+1\qquad  \label{ni+nj ent} \, ;
\nonumber
\\
1 &\leq &n_{i}n_{j}\leq \frac{\left( k+1\right) ^{2}}{4}\, .
\label{ninj}
\end{eqnarray}
One of the eigenvalues of the density operator 
obtained by the partial transposition of the operator
(\ref{3.14}) reads
\begin{equation}
\lambda =\frac{\left| \gamma \right| }{2k}\left( \left| \gamma \right|
\left( n_{i}+n_{j}-2\right) -\sqrt{4|\alpha |^{2}k+|\gamma
|^{2}(n_{i}-n_{j})^{2}}\right) .
\nonumber
\end{equation}
In the non-trivial case of $\left| \gamma \right| >0$ we need only to show 
that 
\begin{equation}
|\gamma |^{2}\left( n_{i}+n_{j}-2\right) ^{2}<4|\alpha |^{2}k+|\gamma
|^{2}(n_{i}-n_{j})^{2}.
\end{equation}
If we use  the 
inequalities (\ref{ni+nj ent}), we find the following constraints 
\begin{eqnarray}
|\gamma |^{2}\left( n_{i}+n_{j}-2\right) ^{2}<|\gamma |^{2}k^{2}\leq
4|\alpha |^{2}k \, ,
\nonumber \\
4|\alpha |^{2}k\leq 4|\alpha |^{2}k+|\gamma |^{2}(n_{i}-n_{j})^{2},
\end{eqnarray}
from which it follows that if 
the condition 
\begin{equation}
0<|\gamma |^{2}k\leq 4|\alpha |^{2}\, ,  \label{cond2}
\end{equation}
is fulfilled then a specific pair qubits described by the density operator
(\ref{3.14}) is entangled.

Till now we have proved that a specific pair of qubits in multi-partite
system is entangled. In order, to show that the corresponding state vector
indeed is associated with a desired entangled graph we have to show that all
other pairs of qubits are separable. Density operators for pairs of
qubits 
$\{i,j\}\notin S$ are given by the expression:
\begin{eqnarray}
\rho _{ij}&=&\\
&&\left( 
\begin{array}{cccc}
|\alpha |^{2}+|\gamma |^{2}\frac{k-n_{i}-n_{j}}{k} & 0 & 0 & 0 \\ 
0 & |\gamma |^{2}\frac{n_{i}}{k} & |\gamma |^{2}\frac{n_{ij}}{k} & 0 \\ 
0 & |\gamma |^{2}\frac{n_{ij}}{k} & |\gamma |^{2}\frac{n_{j}}{k} & 0 \\ 
0 & 0 & 0 & |\beta |^{2}
\end{array}
\right)  \label{MaticaBez}
\nonumber
\end{eqnarray}
with the involved parameters satisfying the
set of inequalities 
\begin{eqnarray}
0 &\leq &n_{i}\leq k  \, ;
\nonumber
\\
0 &\leq &n_{ij}\leq \frac{k}{2} \, ;  \label{nij2} 
\nonumber\\
0 &\leq &n_{i}+n_{j}\leq k\qquad  \label{ni+nj2} \, ;
\nonumber\\
0 &\leq &n_{i}n_{j}\leq \frac{k^{2}}{4}\, .  \label{ninj2}
\end{eqnarray}

Instead of checking that all the eigenvalues of the 
corresponding partially transposed
 operator are
non-negative, we will show that under certain conditions the concurrence of
the state (\ref{MaticaBez}) will be zero.
The eigenvalues of the operator $R$ given by Eq.~(\ref{R}) are 
\begin{eqnarray}
\lambda _{1} &=&\lambda _{2}=|\alpha \beta |^{2}+|\gamma |^{2}\frac{%
k-n_{i}-n_{j}}{k}\, ; 
\\
\lambda _{3,4} &=&|\gamma |^{4}\left( \frac{n_{ij}\pm \sqrt{n_{i}n_{j}}}{k}%
\right) ^{2}\, ;\qquad \lambda _{4}\geq \lambda _{3}\, ,
\nonumber
\end{eqnarray}
and, according to the definition of the 
concurrence (\ref{C}), it is enough to
show that $\lambda _{1}\geq \lambda _{4}$ 
(since then $\lambda _{1}$ is the maximal eigenvalue 
and already $\sqrt{%
\lambda _{1}}-\sqrt{\lambda _{2}}=0$ and so the concurrence vanishes). 
That is, we require that  
\begin{equation}
|\alpha \beta |^{2}+|\gamma |^{2}\frac{k-n_{i}-n_{j}}{k}\geq |\gamma
|^{4}\left( \frac{n_{ij}+\sqrt{n_{i}n_{j}}}{k}\right) ^{2}.
\end{equation}
When we use the inequalities (\ref{ninj2})
we obtain the final condition 
\begin{equation}
|\alpha \beta |^{2}\geq |\gamma |^{4}>0\, ,  \label{cond1}
\end{equation}
which guarantees that the state (\ref{MaticaBez}) is separable.

One can check that there are many states which fulfill the  conditions
(\ref{cond2}) and (\ref{cond1}). In particular, let us assume the state
(\ref{3.12}) with 
\begin{eqnarray}
\alpha &=&\frac{k}{\sqrt{k^{2}+2k+4}} \, ;
\nonumber
\\
\beta &=&\frac{2\alpha }{k} \, ;
\nonumber
\\
\gamma &=&\alpha \sqrt{\frac{2}{k}}\, .
\end{eqnarray}
This state indeed corresponds to the desired graph. This proves that 
one can associate with  an arbitrary entangled graph  a pure
state. Moreover, by construction we have proved that in general 
this state is a superposition of at most $N^2$ vectors from the
$2^N$-dimensional Hilbert space of $N$ qubits. 

\section{Conclusion}

We have introduced a concept of the 
\textit{entangled graphs}: that is an entangled multi-qubit structure
such that 
every qubit is represented by a vertex 
while entanglement between two qubits is represented as an edge between 
relevant vertices.
We have shown
that for every possible graph with non-weighted (see below) 
edges there exists a {\em pure}
state, which represents the graph. Moreover, such state can be constructed
as a superposition of small number of states from
a  subspace of the Hilbert space. The dimension of this subspace grows
linearly with the number of entangled pairs
(thus, in the 
worst case, quadratically with the number of particles).

It is clear that introducing a ``weight'' to the edges of 
the graphs would lead 
to new interesting questions. The weight should correspond to a 
value of the concurrence between the two qubits that are connected by
the given edge.
D\"{u}r in Ref.~\cite{Dur} 
has addressed this question briefly in the context of mixed states
associated with entanglement molecules. As shown in our paper,
the issue of shared bi-partite entanglement in {\em pure} multi-qubit systems
is much more complex issue. Nevertheless, it is of great interest to find
out some general bounds on the amount of possible shared bipartite
entanglement in a given entangled graph. 
Another problem directly related to the issue of weighted graphs is  how 
to maximize bi-partite entanglement for given graphs. 
There, for every graph one could find the optimal state with maximal 
concurrencies on defined pairs of qubits, 
as it was made for specific cases in Refs.~\cite{Chains,Koashi}.

States with defined bipartite entanglement properties are of a possible 
practical use: In communications protocols, like quantum secret sharing 
\cite{Hillery99} or quantum oblivious transfer \cite{Macha} 
one needs many-particle 
states with specific bipartite entanglement properties. 
Therefore deep understanding
 of possible entangled graphs can
help us to understand structure of quantum correlation and the corresponding
bounds  on quantum communications and quantum information processing.

\acknowledgements
We thank M\'{a}rio Ziman and Jakub M\'{a}cha for many helpful discussions.
This work was supported by the IST-FET-QIPC project EQUIP under the contract
IST-1999-11053. VB would like to acknowledge the SFI E.T.S.Walton award.

\end{multicols}


\bigskip

\begin{references}

\bibitem{Schroedinger}
    E.~Schr\"{o}dinger,
       Naturwissenschaften {\bf 23}, 807,   (1935);
 {\it ibid.} {\bf 23}, 823 (1935);
 {\it ibid.} {\bf 23}, 844 (1935).


\bibitem{Peres}
     A.~Einstein, B.~Podolsky, and N.~Rosen,
        Phys. Rev. A {\bf 47}, 777 (1935);
     J.S.Bell,
        Physics {\bf 1}, 195 (1964);
     A. Peres,
        {\it Quantum Theory: Concepts and Methods}
        (Kluwer, Dordrecht, 1993).

\bibitem{Preskill}
     J.~Gruska,
        {\it Quantum Computing}
        (McGraw-Hill,1999);
     J.~Preskill,
         {\it Quantum   Theory   Information and} {\it  Computation}
         ({\tt www.theory.caltech.edu/people/preskill}).

\bibitem{Nielsen}
M.\ A.\ Nielsen and I.\ L.\ Chuang,
{\em Quantum Computation and Quantum Information}
(Cambridge University Press, Cambridge, 2000).


\bibitem{Bennett93}
     C.~H.~Bennett, {\it et al.},
        {Phys. Rev. Lett.} {\bf 70}, 1895 (1993).


\bibitem{Bennett92}
    C.~H.~Bennett and S.~Wiesner,
         {Phys. Rev. Lett.} {\bf 69}, 2881 (1992).

\bibitem{Ekert91}
    A.~K.~Ekert,
         {Phys. Rev. Lett.} {\bf 67}, 661 (1991).

\bibitem{Hillery99}
    M.~Hillery, V.~Bu\v{z}ek, and A.~Berthiaume
          {Phys. Rev. A} {\bf 59}, 1829 (1999);
R. Cleve, D. Gottesman, H. Lo, Phys. Rev. Lett. {\bf 83}, 1874 (1999).

\bibitem{Peres96}
   A.~Peres, Phys. Rev. Lett. {\bf 77}, 4524 (1996).


\bibitem{Horodecki96}
M.~Horodecki, P.~Horodecki, and R.~Horodecki,
Phys. Lett. A {\bf 223}, 1 (1996).


\bibitem{Vedral97}
V.~Vedral, M.B.~Plenio, M.A.~Rippin, and { P.L.Knight},
Phys. Rev. Lett. {\bf 78}, 2275 (1997).


\bibitem{Horodecki2001} 
G. Alber, T. Beth, M. Horodecki, P. Horodecki, R. Horodecki, M R\"{o}tteler,
H. Winfurter, R.F. Werner, A. Zeilinger, Quantum information, Berlin (2001)


\bibitem{CKW}  V. Coffman, J. Kundu, W. K. Wootters, Phys. Rev. A {\bf 61},
052306 (2000)

\bibitem{Chains}  
K.M.~O'Connor and W.K.~Wootters, Phys. Rev. A {\bf 63}, 052302 (2001).

\bibitem{Koashi} M. Koashi, V. Bu\v{z}ek, N. Imoto, Phys. Rev. A {\bf 62}, 
050302 (2000)

\bibitem{note}
This term should not be connected with a totally different
concept of  {\em graph codes} as introduced by 
D. Schlingemann and R.F. Werner
in  Phys. Rev. A {\bf 65}, 012308 (2001)

\bibitem{Dur}  W. D\"{u}r, Phys. Rev. A {\bf 63}, 020303(R) (2001)


\bibitem{Konkurencia}  S. Hill, W.K. Wootters, Phys. Rev. Lett. {\bf 78},
5022 (1997); W.K. Wootters, Phys. Rev. Lett. {\bf 80}, 2245 (1998)


\bibitem{3qubit}  W. D\"{u}r, G. Vidal, J. I. Cirac,
arXiv quant-ph/0005115 (2002).

\bibitem{Macha} J. M\'{a}cha, 
arXiv quant-ph/0005115 (2000)


\end{references}
\end{document}